# Bis(stearoyl) Sulfide: A Stable, Odor-free Sulfur Precursor for High-Efficiency Metal Sulfide Quantum Dot Photovoltaics

*Miguel Albaladejo-Siguan[1,2,3], Anatol Prudnikau[3], Alina Senina[3], Elizabeth C. Baird[1,2,3], Yvonne J. Hofstetter[1,2,3], Julius Brunner[1,2,3], Juanzi Shi,[1,2,3] Yana Vaynzof[1,2,3]\*, Fabian Paulus[3]\**

M. Albaladejo-Siguan, E.C. Baird, Y.J. Hofstetter, J. Brunner, J. Shi, Y. Vaynzof

[1] Institute for Applied Physics and Photonic Materials (IAPP), Technische Universität Dresden, 01187 Dresden, Germany

M. Albaladejo-Siguan, E.C. Baird, Y.J. Hofstetter, J. Brunner, J. Shi, Y. Vaynzof

[2] Leibniz-Institute for Solid State and Materials Research Dresden (IFW), Helmholtzstraße 20, 01069 Dresden, Germany

M. Albaladejo-Siguan, A. Prudnikau, A. Senina, E.C. Baird, Y.J. Hofstetter, J. Brunner, J Shi, Y. Vaynzof, F. Paulus,

[3] Center for Advancing Electronics Dresden (cfaed), Technische Universität Dresden, 01069 Dresden, Germany

E-mail: yana.vaynzof@tu-dresden.de, fabian.paulus@tu-dresden.de

**Abstract:**

The synthesis of metal sulfide nanocrystals is a crucial step in the fabrication of quantum dot (QD) photovoltaics. Control over the quantum dot size during synthesis allows for precise tuning of their optical and electronic properties, making them an appealing choice for electronic applications. This flexibility has led to the implementation of quantum dots in both highly-efficient single junction solar cells and other optoelectronic devices including photodetectors and transistors. Most commonly, metal sulfide quantum dots are synthesized using the hot-injection method utilizing toxic, and air- and moisture sensitive sulfur source: bis(trimethylsilyl) sulfide $(TMS)_2S$. Here, we present bis(stearoyl) sulfide ($St_2S$) as a new type of air-stable sulfur precursor for the synthesis of sulfide-based QDs, which yields uniform, pure, and stable nanocrystals. Photovoltaic devices based on these QDs are equally efficient as those fabricated by $(TMS)_2S$ but exhibit an enhanced operational stability. These results highlight that St$_2$S can be widely adopted for the synthesis of metal sulfide quantum dots for a range of optoelectronic applications.

## 1. Introduction

Colloidal metal sulfide nanocrystals are a well-studied class of low-dimensional semiconductors that have found applications in a variety of optoelectronic devices such as solar cells (SCs),[1,2] photodetectors (PDs),[3,4] light-emitting diodes (LEDs),[5,6] field-effect transistors (FETs),[7,8] and lasers.[9,10] They are processed as colloids with a functionalized surface, making it easy to manipulate their properties on a large scale as a low-cost semiconducting material.[11–13] In the case of nanocrystals with size smaller than twice the exciton Bohr radius, quantum confinement effects determine their electronic structure, making their properties fundamentally different to their bulk counterparts.[14] Properties like a size-dependent bandgap,[15] surface-dictated charge transport,[16] collective phenomena of superstructures,[17,18] and strong infrared absorbance,[19] are all derived from the quantum confinement effect in these nanometer-sized particles, which are usually referred to as quantum dots (QDs).

According to their elemental composition, metal sulfide QDs are classified as binary (e.g. PbS, Ag$_2$S), ternary (e.g. AgBiS$_2$, CuInS$_2$), or quaternary (e.g. Cu$_2$ZnSnS$_4$, Ag$_2$ZnSnS$_4$) semiconductor compounds. Since about 50% of the solar spectrum lies in the infrared region,[20] narrow-bandgap metal sulfide QDs encompassing both the visible and near-infrared parts of the electromagnetic spectrum are of particular interest for solar cell applications. Out of the different metal sulfide QDs, binary PbS QD based SCs have gathered most attention due to their high performance, facile and low-cost synthesis and processing, and encouraging possibilities for large-scale fabrication. Recently, a record power conversion efficiency (PCE) of 15.45% has been achieved for PbS QD SCs based on extensive device engineering with optimized extraction layers.[21] The latest advances for these type of devices are focused on either the optimization of the extraction layers or the improvement of interdot coupling.[22–24]

Heavy metal-free sulfide-based quantum dots have recently attracted significant attention as a more environmentally-friendly alternative to PbS. In particular, ternary $AgBiS_2$ QDs are a promising candidate as a low-toxicity absorber material in photovoltaic cells.[25] $AgBiS_2$ QDs exhibit ideal properties for photovoltaics with a bandgap of ca. 1.0-1.1eV, very high absorption coefficient and sufficiently good stability.[26] As a result of improving device structure and synthetic approaches, solar cells based on $AgBiS_2$ QDs with PCE of over 9 % have recently been demonstrated.[27,28]

The procedure for PbS nanocrystal synthesis, which is used for almost all high-performing PbS QD photovoltaic devices, follows a protocol first proposed by Margaret Hines in 2003.[29] Hines' method uses lead oleate and bis(trimethylsilyl) sulfide ($(TMS)_2S$) as reagents, and is often adapted with only minimal modifications in temperature, reactant ratios, and purification steps. As in the case of PbS QDs, high performance SCs have been demonstrated only for $AgBiS_2$ QDs obtained using $(TMS)_2S$ as a sulfur precursor. The use of $(TMS)_2S$ as a sulfur precursor has many advantages, such as a small size dispersity, good reactivity and high chemical yield of the final product.[29–32] Despite these beneficial properties, $(TMS)_2S$ as a QD precursor suffers from several intractable disadvantages. One major drawback of the synthesis is the low reproducibility, especially regarding the final QD size and stability of the colloids. For example, even when employing the same synthesis conditions, the position of the first excitonic peak may shift by as much as 150 nm simply due to evolution of the properties of the $(TMS)_2S$ precursor over time. This is mostly attributed to the instability of $(TMS)_2S$, which hydrolyzes to $H_2S$ in the presence of water and can decompose quickly upon exposure to the ambient or simply by storage in non-inert conditions.[33] In addition, $(TMS)_2S$ is a flammable liquid with a ghastly smell and a relatively high price[11] which might further hamper attempts to upscale the synthetic procedures for large-scale QD fabrication. Several studies have therefore investigated alternatives for $(TMS)_2S$ as sulfur precursor, some of which even offer lower costs for the resulting product, by employing cheaper starting materials. Among the most promising candidates is thiourea,[34] which is also toxic and even suspected to cause cancer.[35] A derivative of thiourea, N,N-diphenylthiourea, was used by Ma and co-workers to fabricate directly lead halide capped PbS QDs for solar cell applications in a low-temperature process, providing a low-cost access for PbS.[36] Other attempts to replace (TMS)$_2$S utilized elemental sulfur in amines that decomposes via polysulfide intermediates in-situ to H$_2$S.[37,38] Yuan et al. used an ionic liquid formed upon introduction of H$_2$S in oleylamine as precursor, that provides a cost-effective path for the synthesis of PbS.[39]

Elemental sulfur has been used to synthesize highly monodisperse PbS QDs (size dispersity <5%) with core sizes ranging from 4.3 to 8.4 nm.[40,41] However, Weidman et al. demonstrate in their study, obtaining the size that is relevant for single junction photovoltaic applications (around 3 nm for 1.3 eV) involves reducing the synthesis temperature to 40°C and thus increasing the size dispersity and affecting the consistency of the nanocrystal synthesis.[40] The size range using elemental sulfur can be extended by the addition of tri-n-octylphosphine to the synthesis (dot size 3-10 nm), but the size dispersity is then increased to 10%.[42] The performance of PbS QD solar cells made with elemental sulfur has not yet surpassed the efficiencies of (TMS)$_2$S based synthetic protocols, due to differences in surface passivation of these QDs. To the best of our knowledge, optoelectronic devices with sulfide-based nanocrystals synthesized from elemental sulfur have not overcome the performance of (TMS)$_2$S devices. In 2014, Yuan et al. demonstrated solar cells with PbS QDs with 1 eV bandgap, which were made using elemental sulfur, and reached only a 5.4 % power conversion efficiency.[43] A similar synthesis has been recently performed in the case of AgBiS$_2$, and the photovoltaic performance reaches 5.5%.[44] For PbS nanocrystal field-effect transistors, the addition of elemental sulfur is shown to modify the shape and size of the dots, as well as introducing p-type doping.[45]

Here we propose an alternative sulfur precursor for the synthesis of metal sulfide QDs for photovoltaic applications: bis(stearoyl) sulfide (St$_2$S). This precursor can easily be synthesized under laboratory conditions and has already been successfully applied for the synthesis of $CdS_xSe_{1-x}$ nanoplatelets.[46] Its solid form at room temperature simplifies storage and handling and provides a prolonged lifetime in ambient conditions without any signs of degradation, thus resolving the irreproducibility issues associated with the instabilities of $(TMS)_2S$. With a melting point of 65°C and good solubility in nonpolar organic solvents,[47] it can easily be applied in hot-injection synthesis techniques to produce monodisperse nanocrystal dispersions. We use the $St_2S$ precursor to synthesize sulfide-based QDs such as PbS and $AgBiS_2$ and demonstrate that they exhibit similar morphological and compositional characteristics as those synthesized by $(TMS)_2S$. Furthermore, PbS solar cells made using $St_2S$-QDs exhibit similar performance but higher stability to those fabricated by $(TMS)_2S$, even without further optimization.

## 2. Results and Discussion
### 2.1 Synthesis of metal sulfide nanocrystals

Sulfide-based quantum dots were synthesized by adopting the previously reported procedures for PbS and $AgBiS_2$ record-efficiency solar cells.[21,27] These approaches include the injection of $(TMS)_2S$ into a hot solution of the corresponding metal oleates at a temperature of 100 °C in an inert atmosphere. $(TMS)_2S$ and $St_2S$ were used as sulfur precursors in these syntheses. **Figure 1** shows the chemical structure of the two sulfur-containing precursors. For both $St_2S$ and $(TMS)_2S$ precursors, sulfur is already present in the required oxidation state (-2), which ensures a fast reaction with metal carboxylates upon addition to the reaction mixture. This is different to the synthesis using elemental sulfur and oleylamine, where sulfur has to be reduced in advance, typically by the application of heat.[41] Several studies have shown that this leads to a complex mixture of species, especially polysulfides ($S_4^{2-}$, $S_6^{2-}$) and these preceding equilibriums can impact the quantum dot synthesis.[38]

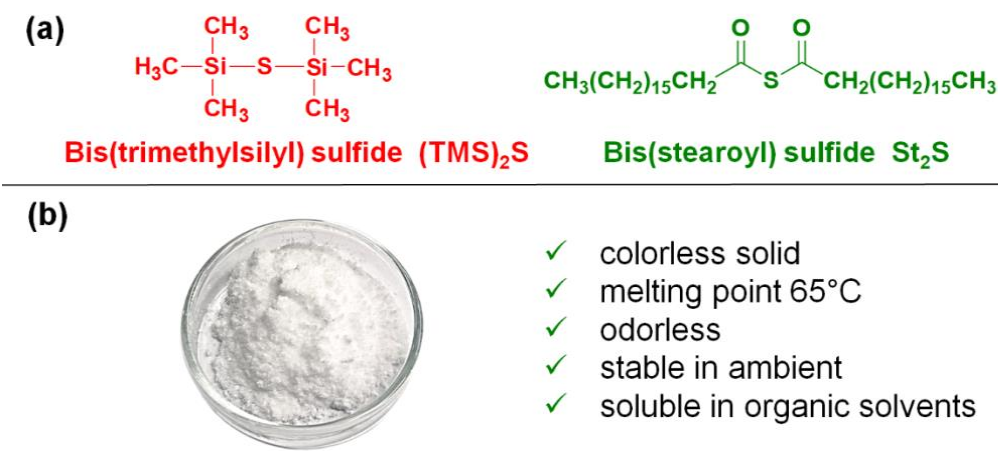

**Figure 1.** (a) Molecular structure of the sulfur precursors used in this work (a) and appearance and material properties of St$_2$S.

Bis(stearoyl) sulfide was synthesized using the multi-step one-pot method developed by Koketsu *et al.*.[47] The synthesis involves the reaction of stearoyl sulfide with LiAlHSH obtained by reducing elemental sulfur with lithium aluminum hydride (LiAlH$_4$) in tetrahydrofuran. After careful purification (see experimental methods section for details), St$_2$S is obtained as colorless powder, soluble in non-polar organic solvents such as toluene or chloroform. The synthesized precursor was characterized by $^1$H and $^{13}$C nuclear magnetic resonance (NMR) spectroscopy (Figure S1). Purified St$_2$S is odorless and can be stored under ambient conditions for several months, since bis(acyl) sulfides are the most stable of all chalcogencarboxylic acid anhydrides.[48] In contrast, (TMS)$_2$S is a foul-smelling liquid that readily hydrolyses upon contact with air, making it challenging to reliably use in routine laboratory research. Moreover, we compared the costs for commercially available (TMS)$_2$S from various suppliers with the laboratory synthesis of St$_2$S. Even though only a laboratory scale synthesis is employed, the costs of 1 mmol St$_2$S are comparable with the price of a commercially (TMS)$_2$S (see Tables S1, S2, and S3). Moreover, St$_2$S does not require special storage and transportation conditions like the unstable, toxic and flammable (TMS)$_2$S, which might facilitate an easier commercializing of the proposed precursor. After the injection of a 1-octadecene (ODE) solution of (TMS)$_2$S into a hot lead oleate solution, the precursor immediately reacts with the metal oleate to form PbS nanocrystals. In contrast, after the injection of the toluene solution of St$_2$S into lead oleate solution at the same temperature, the reaction commenced only after a subsequent injection of a small amount of oleylamine, which results in the formation of PbS nanocrystals. We propose that the amino group of the oleylamine acts as a nucleophile which attacks the carbonyl centers of St$_2$S, resulting in a fast St$_2$S decomposition to a reactive sulfur species. That allows the reactivity of St$_2$S to be tuned by changing the ratio between precursor and the amount of nucleophile added, giving an additional parameter to control the synthesis and consequently, the PbS quantum dot size (see section 2.2.2). It is also important to note here, that utilizing St$_2$S allows to maintain a reactant ratio Pb:S of 2:1, identical to (TMS)$_2$S. Past studies utilizing elemental sulfur in oleylamine had to employ a much more lead rich ratio Pb:S of 24:1 to obtain small size distributions, discarding most of the lead precursor after synthesis.[40]

In the case of the AgBiS$_2$ QD synthesis, St$_2$S reacts with metal carboxylates as rapidly as (TMS)$_2$S without the presence of oleylamine. The approach we used to synthesize AgBiS$_2$ QDs involves for both precursors a significant excess of oleic acid, which likely provides a suitable chemical environment for the formation of reactive sulfur species.

### 2.2.1 Morphological and Compositional Analysis

Possible differences in the size, morphology and composition of the QDs made, using (TMS)$_2$S and St$_2$S as sulfur precursors were investigated by means of transmission electron microscopy (TEM), optical absorbance spectroscopy (UV-vis) and x-ray diffraction (XRD) measurements. TEM images of PbS QDs (**Figure 2**) show a similar shape for both nanoparticle sets, which is known by past studies to be that of a truncated octahedron.[49] The size of the PbS QDs is ~2.70 nm for (TMS)$_2$S ($\sigma = 0.25$) and ~2.73 nm for St$_2$S ($\sigma = 0.32$) precursors (see Figure S2). These deviations in size and the size distribution are rather small and are within the typically much larger batch-to-batch variation normally observed in (TMS)$_2$S synthesis (Figure S3). As is shown in Figure S3, our experience shows that multiple syntheses performed using (TMS)$_2$S can lead to quantum dots with starkly different sizes, as is evidenced by significant spread in the position of their first excitonic peak measured by UV-vis absorption. This position can vary by as much as 150 nm, despite the use of the same synthesis procedure, which occurs for many different injection temperatures. This illustrates the low level of reproducibility offered by (TMS)$_2$S due to its poor stability.

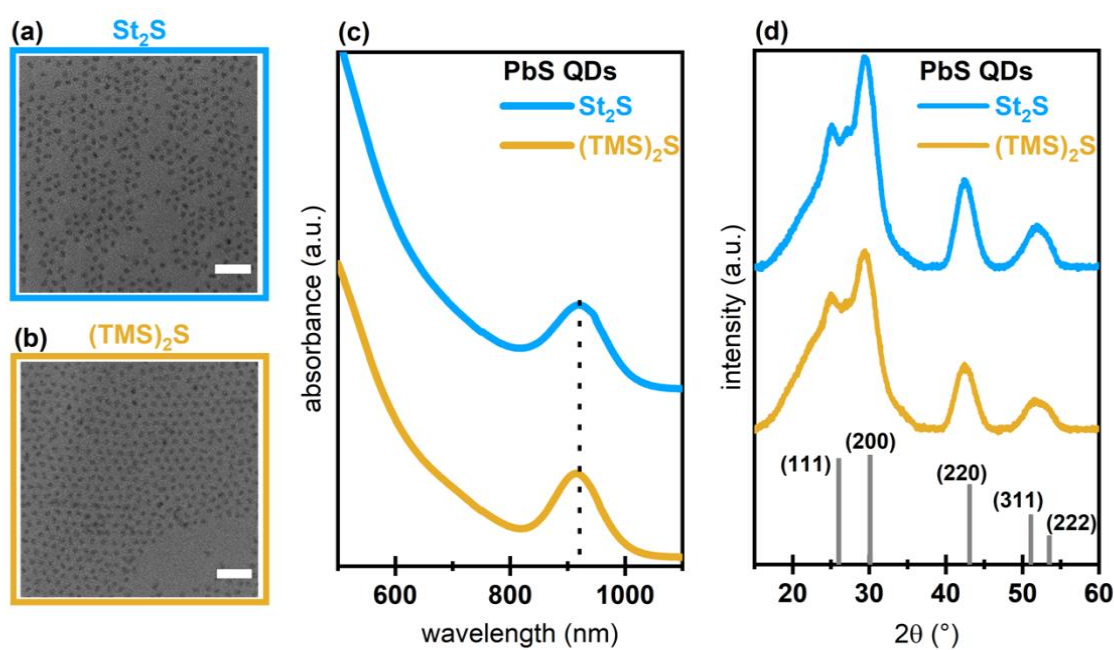

**Figure 2.** TEM images of PbS QDs synthesized with (a) St$_2$S and (b) (TMS)$_2$S as a sulfur precursor. The scale bar is 20 nm. Comparison of the absorbance spectra of differently synthesized PbS QDs in octane (normalized, offset) (c). XRD patterns of PbS QDs (d). The vertical lines are the powder pattern for the bulk reference of PbS (ICSD, code 5533).

The absorbance of the synthesized nanocrystals in solution was measured on a UV-vis spectrometer, and the resulting spectra are presented in **Figure 2c**. The distinct first excitonic peak position, which is related to the size and optical bandgap of the QDs, is different for each PbS QD solution. For the (TMS)$_2$S reference, the peak position of 915 nm was chosen for its optimal bandgap for solar cell applications following the thermodynamic limits.[50] Note that the bandgap onset will redshift after ligand exchange with lead halides to around 1.3 eV.[51] A slightly redshifted and larger width of the peak is observed for St$_2$S QDs, which agrees well the minimally larger particles in case of St$_2$S, which translates to a difference of 12 meV for their respective bandgaps. To evaluate the impact of sulfur precursor on the optical properties of the synthesized quantum dots we conducted photoluminescence (PL) measurements on these colloids in octane (Figure S4). The emission peak of the PbS QDs centers around 1000 nm in both cases and appears to be slightly broader in the case of St$_2$S-based QDs, which could reflect the slightly broader size distribution obtained from synthesis. Time-resolved PL measurements, however, reveal that there is no substantial difference in the temporal decay of the excited state (Figure S5). From this we conclude that St$_2$S-based QDs exhibit similar properties and quality to (TMS)$_2$S based QDs and are not more defective.

The nanocrystal structure was analyzed by XRD on drop-cast, micrometer thick films of PbS QDs. As shown in **Figure 2d**, the XRD diffractograms of the nanoparticles resemble those of the bulk material, but with broader reflections. This is characteristic for nanometer sized QDs,[52] and agrees well with previous reports examining PbS QDs of similar size.[53] No differences in peak position and peak width are observed for PbS QDs when comparing (TMS)$_2$S and St$_2$S precursors, which confirms a very similar crystalline structure and size of the QDs obtained with each sulfur precursors. Compositional analysis of the PbS QDs was performed via x-ray photoemission spectroscopy (XPS) on thin films with PbS QDs with a Pb-halide ligand shell. Pb and S spectra are plotted in **Figure 3**, with neither showing significant changes dependent on the sulfur precursor used in QD synthesis. No metallic Pb is observed in either sample, indicating the high quality of the PbS QDs. The atomic percentages of the different elements are shown in Table S4. The Pb:S ratios for St$_2$S and (TMS)$_2$S QDs are very similar (1.22 and 1.27, respectively) whose small variation might be caused by different coverage by the lead containing ligands. The halide composition of the ligand shell is shown in Figure S6, where the XPS spectra of iodine and bromine are shown. Both PbS QD films exhibit a I/Br ratio of ~4.3, as calculated by the XPS fits, which corresponds well to the initial halide ratio employed in ligand exchange.[22,54] Interestingly, the oxygen content in the (TMS)$_2$S QD film is almost 50% higher than that in the St$_2$S QD film, which might indicate they are less prone to oxidation upon exposure to the ambient.

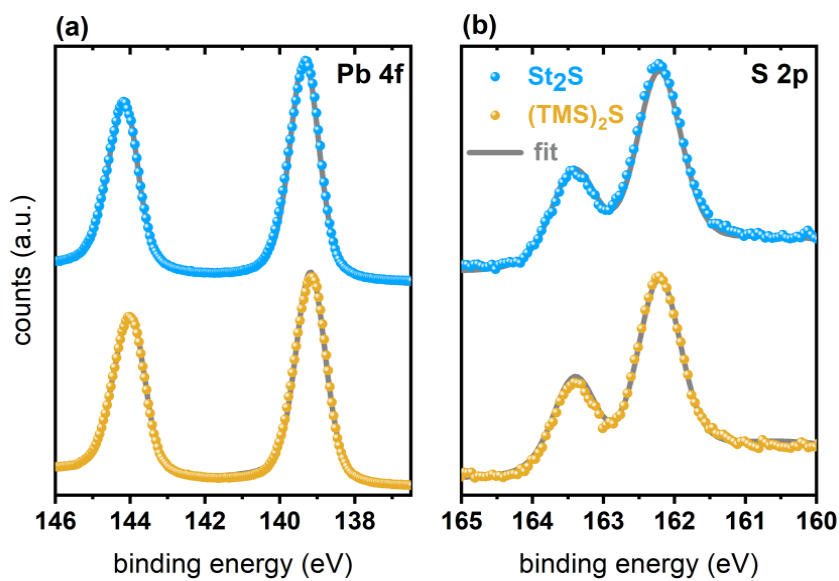

**Figure 3.** XPS spectra of PbS QDs with lead halide ligands. The (a) Pb 4f and (b) S 2p signals are similar for the reference (TMS)$_2$S and modified St$_2$S synthesis.

While the structural, optical and compositional properties of PbS remain largely unaffected by the choice of sulfur precursor, the properties of AgBiS$_2$ QDs are much more influenced by the choice of sulfur precursor. TEM imaging of the QDs synthesized using St$_2$S reveals significantly smaller QDs than those made with (TMS)$_2$S, with an average size of 3.8 nm compared to 4.5 nm despite quasi-identical synthesis conditions (Figure 4). Employing St$_2$S as sulfur precursor, however, leads to a much more uniform QD size and a narrower size distribution (Figure S7). The different QDs sizes are also evidenced by the difference of the absorbance onset, which consequently shows a larger bandgap for smaller AgBiS$_2$-QD synthesized with St$_2$S. The absorbance of (TMS)$_2$S QDs extends further into the near-infrared-region with a less pronounced onset. This could be indicative of a higher concentration of larger aggregates in the case of (TMS)$_2$S that are also visible in the TEM images. XRD measurements on drop-cast films confirm the above-made observations and show significantly broader reflections for the smaller St$_2$S-QDs. The peak position however reveals that ternary AgBiS$_2$ QDs were produced in both cases. The appearance of the unexpected reflection at 20° in the XRD patterns of AgBiS$_2$ QDs using St$_2$S as precursors can be attributed to the packing of alkyl chains on the nanocrystal surface.[55] The use of aliphatic, saturated St$_2$S as a sulfur precursor can cause the presence of stearate anions as ligands on the nanocrystal surface. It is known that such long-chained ligands with fully saturated alkyl chains, can pack better on the nanocrystal surface than the commonly employed, unsaturated, bent oleates.[56] To test this assumption, we treated the St$_2$S-based AgBiS$_2$ QDs with an excess of oleic acid. As shown in Figure S8, the peak at 20° vanishes after the treatment of QDs with oleic acid, indicating that this signal on the XRD pattern is the result of the presence of stearate species in the system. In addition, the introduction of 2 mmol of steric acid (equal to 1 mmol of St$_2$S in terms of stearate species amount) to the synthesis of AgBiS$_2$ QDs using (TMS)$_2$S also results in the presence of a peak at 20° in the XRD pattern (Figure S9).

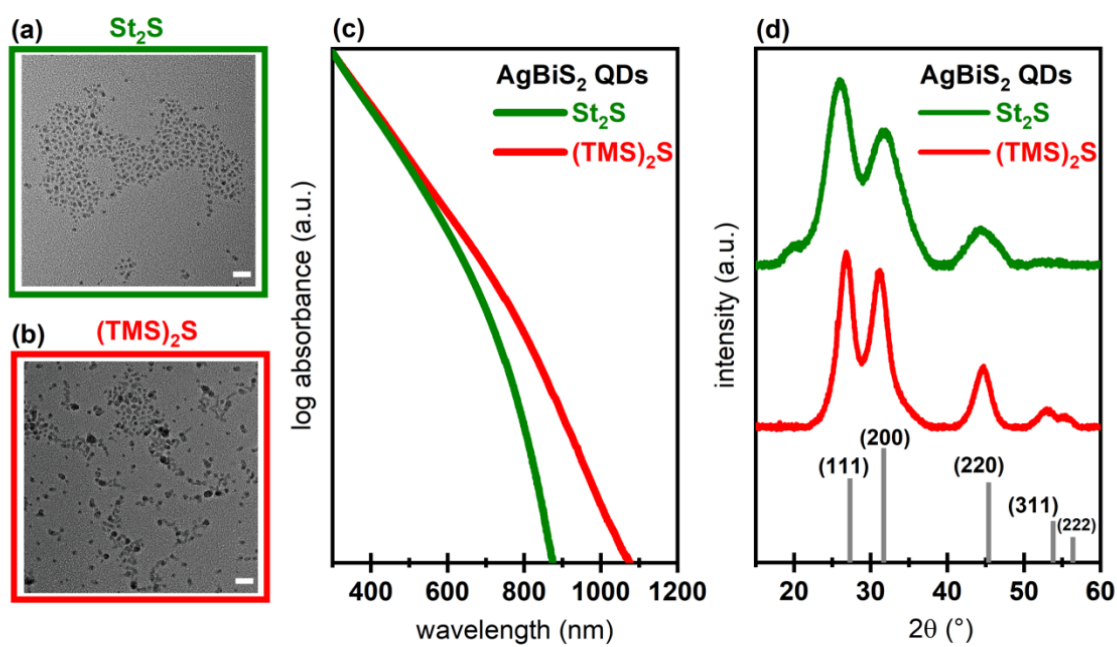

**Figure 4.** TEM images of AgBiS$_2$ QDs synthesized with (a) St$_2$S and (b) (TMS)$_2$S as a sulfur precursor. Scale bar is 20 nm. (c) Absorbance spectra of AgBiS$_2$ QDs in toluene normalized to absorbance at 300 nm, (in logarithmic scale showing three orders of magnitude). (d) XRD patterns of AgBiS$_2$ drop-cast QDs films. The vertical lines are the bulk reference of AgBiS$_2$ (ICSD, code 604845).

The composition of AgBiS$_2$ is strongly impacted by the choice of sulfur precursor, as shown by XPS measurements on thin films after solid phase ligand exchange with tetramethylammonium iodide (TMAI) in methanol. These were produced in a layer by-layer deposition process as described in literature.[25] This is particularly evident in the Bi 4f spectra (Figure 5a, Table S5), where the (TMS)$_2$S samples result in two different Bi species, with a Bi 4f$_{7/5}$ peak at binding energies of 158.1 and 158.8 eV. The presence of two Bi species is commonly reported for QDs fabricated using (TMS)$_2$S,[25,57] with the low binding energy species associated with Bi AgBiS$_2$

QDs, while the high binding energy peak was suggested to be associated with Bi-I bonds formed during ligand exchange.[25] However, our previous studies examining the degradation behavior of $AgBiS_2$ showed that the high binding energy peak is associated with $Bi_2O_3$, since upon exposure to oxygen of the QD films, its relative contribution to the Bi spectrum grew substantially.[58] This assignment to $Bi_2O_3$ is further corroborated by the presence of a strong peak at 529.6 eV in the O 1s spectra measured on the $(TMS)_2S$ samples, which is consistent with a metal oxide (Figure 5c). On the other hand, the synthesis with $St_2S$ results in a single Bi species at 158.1 eV, indicating that no $Bi_2O_3$ is formed. This suggests that the $AgBiS_2$ QDs fabricated using $St_2S$ are far more stable than those made using $(TMS)_2S$.

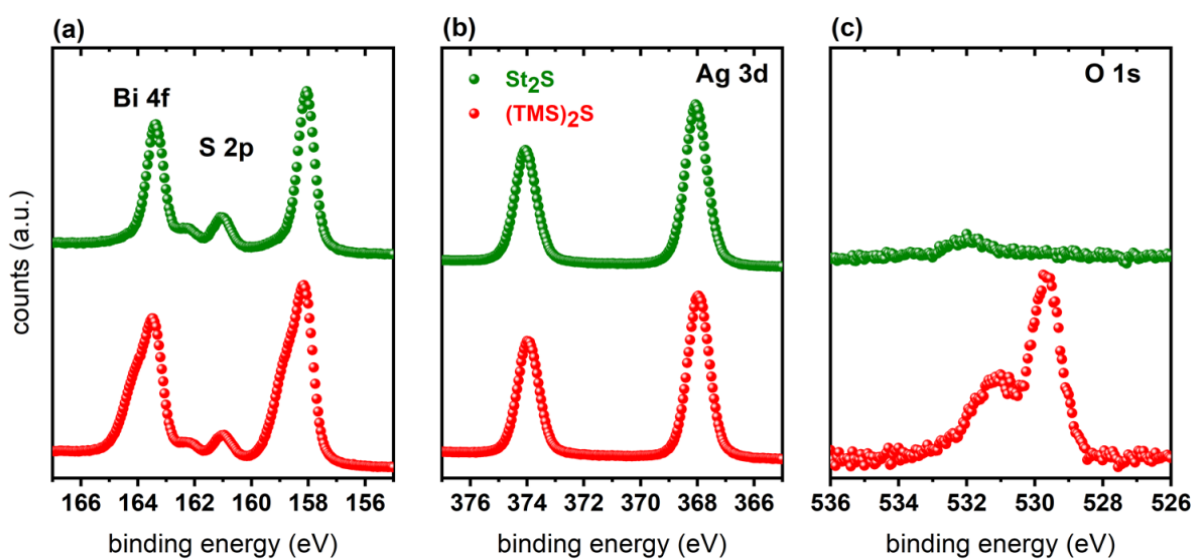

**Figure 5:** XPS spectra of $AgBiS_2$ nanocrystals after TMAI ligand exchange. The (a) Bi 4f shows a clear shoulder at higher binding energies in case of $(TMS)_2S$ reference while the (b) Ag 3d doublet remains unaffected by the choice of precursor.

### 2.2.2 Size tunability using $St_2S$

Modifying the size of colloidal QDs is an essential requirement for any new precursor to adjust the QD optoelectronic properties. To demonstrate the feasibility of size-tuning with $St_2S$ we conducted several experimental series. The size of PbS QDs can be tuned over a wide range by varying the reaction temperature, time, concentration and order of injection as shown in Figure S10 and Table S6, allowing to produce PbS QDs of various sizes, with an absorbance wavelength position of the first excitonic peak ranging from 880 nm to 1525 nm. Even for large

PbS QDs a narrow size distribution of ca. 10% size dispersity can be achieved, as demonstrated in Figure S11 for PbS QDs (size 5.1 nm) with an 1$^{st}$ excitonic peak centered at 1400 nm. Since oleylamine is required as a nucleophile to trigger decomposition of St$_2$S in the reaction system for PbS, modifying its overall concentration allows further tuning of the QD size (Figure S12). Higher amounts of oleylamine result in smaller PbS QDs, evidenced by a blue shift and broadening of the 1st excitonic peak as compared to small amounts of oleylamine. We assign this effect to result from an increased reactivity of St$_2$S in the presence of larger quantities of oleylamine, which leads to the consumption of more precursor at the nucleation stage and, consequently, a lower contribution of the growth stage to the final size of the nanocrystals, resulting overall smaller nanocrystals. Using a high volume of oleylamine (8 mL) results in a longer injection time and a large number of nucleation events and hence a broad size distribution (black line, S8). The size of AgBiS$_2$ QDs can also be easily tuned by modifying the synthesis temperature from 100°C to 140°C, allowing to produce QDs ranging from (4.2 ± 0.9) nm to (6.4 ± 1.2) nm (see Table S7 and Figure S13 and S14). Our results demonstrate that St$_2$S can be used as a sulfur precursor and allows to produce QDs with a large range of sizes, while oleylamine can be used to fine-tune the size and size distributions. These results highlight that St$_2$S is a viable precursor for QDs with absorption and emission features from the visible to the infrared and is not limited to the synthesis of QDs of a particular size.

**2.4 Photovoltaic Device Performance**

Solar cells of all QDs were fabricated by replacing the long oleate ligands from synthesis with shorter ligands that allow sufficient charge transport in the films. For PbS QDs a mixture of lead iodide and lead bromide in a molar ratio of 10:2 was employed, following the liquid phase exchange procedure by Sargent and co-workers.[54] For AgBiS$_2$ QDs ligand exchange is performed using TMAI in a solid-phase ligand exchange.[25] The device architectures of both solar cells are shown in **Figure 6a-b**. Both solar cell architectures employ zinc oxide (ZnO) as an electron extraction layer on transparent indium tin oxide (ITO). The active layer of PbS solar cells is deposited in a single step resulting in layer thickness of ca. 375 nm, followed by a hole extraction layer of PbS-EDT dots and gold contacts. The QDs used for the EDT layer are the same QDs as in the active layer, only treated with EDT as ligand (see experimental section for details). $AgBiS_2$ layers were deposited in a layer-by-layer approach with final layer thickness of ca 35 nm, followed by a layer of doped poly(triaryl amine) (PTAA) and silver electrodes. While we recently demonstrated that PTAA can act as efficient hole extraction layer for $AgBiS_2$ when doped with the fluorinated fullerene $C_{60}F_{48}$,[59] we employed here for the first-time tris(pentafluorophenyl)borane (BCF) as dopant in $AgBiS_2$ solar cells, that has been shown to effectively dope PTAA in perovskite solar cells.[60]

The power conversion efficiencies (PCEs) of the PbS and $AgBiS_2$ QD solar cells, from both $(TMS)_2S$- and $St_2S$-based QDs synthesis, are shown in **Figure 6c-d**, with the full photovoltaic parameters presented in Figure S15. In the case of PbS QD solar cells, both sulfur precursors lead to a similar overall performance with PCEs between 10% and 11%. We note that the open-circuit voltage ($V_{OC}$) is on average slightly higher for the reference $(TMS)_2S$ devices, which is attributed to the aforementioned minimally redshifted bandgap of the $St_2S$ QDs. In fact, this shift in $V_{OC}$ coincides with the shift of the first excitonic peak of 12 meV observed in Figure 2c. While $St_2S$-QD devices exhibit a slightly lower $V_{OC}$, their short-circuit current density ($J_{SC}$) is significantly higher reaching 27.2 mA cm$^{-2}$ as compared to 26.5 mA cm$^{-2}$ for $(TMS)_2S$. This fact is sustained by the external quantum efficiency of the devices, which is higher and redshifted for $St_2S$ solar cells (Figure S16). A similar fill factor (FF) of around 63% results for both types of devices, with a mean power conversion efficiency of 10.6% under one sun illumination. These results show that $St_2S$ can be a suitable replacement for the problematic $(TMS)_2S$ in the synthesis of PbS QDs without any further optimization of the synthetic procedure or the device structure.

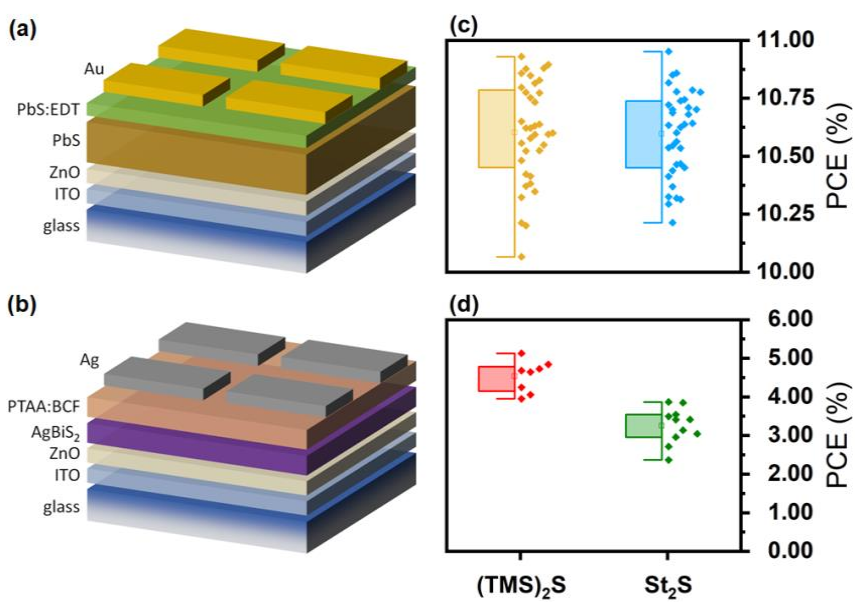

**Figure 6.** Schematic Device architecture of (a) PbS QD solar cells and (b) AgBiS$_2$ solar cells. Solar cell performance parameters of PbS-QD solar cells utilizing different sulfur precursor: (c) PCE of PbS and (d) PCE of AgBiS$_2$.

On the other hand, solar cells based on AgBiS$_2$ QDs exhibit significant differences in performance, depending on the sulfur precursor used. Solar cells with (TMS)$_2$S QDs achieve an average performance of 4.5% PCE (see Figure 6d), while solar cells employing St$_2$S-QDs show a lower performance of approximately 3.5%. This low PCE stems from the devices' significantly lower V$_{OC}$ despite the fact, that these smaller QDs possess a larger bandgap. We associate this with an overall larger defect concentration on their surfaces after the solid-state ligand exchange that negatively impacts the open-circuit voltage (Figure S17). Furthermore, the larger distribution and overall lower average of the short-circuit current density could indicate a non-complete ligand exchange for these small dots, that is also represented in a lower EQE of these devices (Figure S18). This suggests that the synthesis of AgBiS$_2$ based on St$_2$S would need to be further optimized and would thus differ from the established procedures for (TMS)$_2$S. Nonetheless, the fact that St$_2$S leads to high-quality AgBiS$_2$ without the presence of Bi$_2$O$_3$ makes this precursor very promising for future studies also for this quantum dot type.

## 2.5 Stability of the devices

Considering the promising indications from the XPS measurements regarding the stability of quantum dots fabricated using St$_2$S, we examined the performance evolution of the higher performing PbS solar cells. Specifically, the devices were continuously characterized by measuring the current density-voltage characteristics for 100h under constant illumination and in a humid air environment (40% relative humidity) at 23 °C, similar to previous stability studies.[61] The evolution of the photovoltaic parameters for devices fabricated using (TMS)$_2$S and St$_2$S are displayed in Figure 7. While both types of devices display similar degradation dynamics, solar cells based on St$_2$S QDs proved to be more stable, retaining 60% of their initial performance, while the reference (TMS)$_2$S cells dropped to 40% of their original performance in the same time span.

The evaluate the origin of this difference in stability we first compared the stability of the QDs after synthesis in dispersion. Over a period of 90 days, we could not observe differences in the stability between the (TMS)$_2$S- and St$_2$S-based QDs (see Figure S19). We therefore focused on the EDT-treated QD films by means of XPS, since the degradation of this layer has been shown to dominate the overall device degradation dynamics.[62] Similarly to the PbX$_2$ treated QDs, the treatment of the QDs with EDT results in a significantly higher amount of oxygen species in the case of (TMS)$_2$S-based QDs compared to St$_2$S-based QDs (Figure S20b). Moreover, the (TMS)$_2$S QDs exhibit higher amounts of unbound and partially oxidized thiolates centered around 164 eV in binding energy. These thiolates are prone to undergo oxidation first and can lead to a much faster degradation of the overall device performance (Figure S20a).[62]

We believe that the choice of different precursors in the synthesis results in different surface properties of the final QDs, that affect the binding properties to ligands. In the case of the (TMS)$_2$S-based synthesis, only oleate anions are assumed to terminate the charged {111} facets of the PbS QDs to provide charge-neutral nanoparticles. However, these oleate anions cannot completely passivate the {111} facets due to the steric hindrance of the bent carbon chains caused by the presence of the C=C double bond.[64] As shown in a study by Zherebetskyy et al. water can result in the formation of hydroxyl groups on the {111} facets of PbS QDs, even in between oleic acid ligands.[65] Such partially a hydroxylated surface in conjunction with physisorbed water through hydrogen bonding results in high oxygen amounts in the XPS spectra. The strong bonding of hydroxyl groups at the surface makes it difficult to replace them and will consequently lead to a less efficient passivation with EDT. In the case of St$_2$S-based PbS QDs, the nanocrystal surface can be passivated by both oleate and stearate anions during synthesis. The presence of a straight, fully saturated ligand will result in a denser ligand shell and a better passivation of the PbS QD surface. Our results show that St$_2$S lowers the overall amount of a hydroxylated PbS QD surface. This is in line with a recent study by Wang et al. in which the authors demonstrated that introducing short-chained acetate anions in the synthesis of oleate-capped PbS QDs provides a more efficient termination of the nanocrystal surface with carboxylates and reduces the overall amount of surface hydroxyl groups.[66] In the post-synthetic ligand exchange process, carboxylates are more easily replaced by EDT than hydroxyl groups, which leads to more effective passivation of the QDs surface and an increased stability of solar cells in the case of St$_2$S.

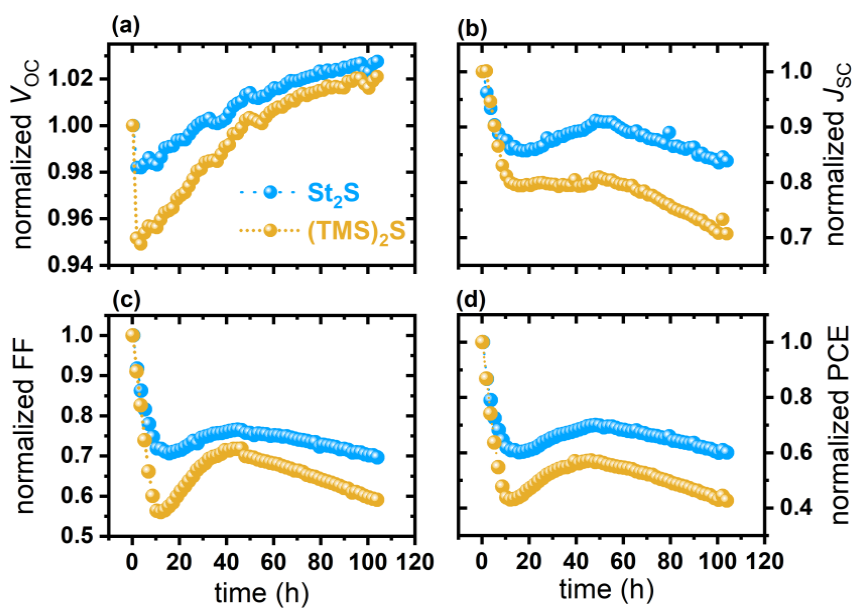

**Figure 7.** Solar cell stability testing under continuous 100 mW cm$^{-2}$ illumination (1 sun) and humid air. Devices made with St$_2$S PbS quantum dots are more stable after 100 h of illumination (65% of initial PCE versus 45 % for (TMS)$_2$S).

## 3. Conclusion

In this study, we have demonstrated that the sulfur precursor (TMS)$_2$S can be effectively replaced by stearoyl sulfide St$_2$S for the synthesis of sulfide-based QDs. We demonstrated that St$_2$S can be readily adopted in the existing synthetic procedures for binary PbS and ternary AgBiS$_2$ QD synthesis, without a dramatic impact on the optoelectronic application of these materials. In the case of binary PbS, the resulting structural, optical and compositional properties of the St$_2$S synthesized QDs are very similar to those made with (TMS)$_2$S, but differ in the case of ternary AgBiS$_2$ QDs. Consequently, the solar cell performance of St$_2$S-PbS-QDs is comparable to the ones obtained by (TMS)$_2$S-PbS-QDs, while their stability is markedly enhanced. In case of ternary AgBiS$_2$ QDs, the use of St$_2$S results in smaller QDs with narrow size distribution and slightly altered chemical composition, without any changes to the synthetic protocol. Compared to larger (TMS)$_2$S QDs the overall solar cell performance is lower, suggesting that further optimization of the synthetic protocols is required to achieve ternary QDs with similar properties, compared to (TMS)$_2$S. Nevertheless, our experiments show that St$_2$S can act as an air-stable and easy-to-handle alternative in the fabrication of sulfide-based QDs for photovoltaic applications and can replace the air-sensitive, foul-smelling and toxic (TMS)$_2$S commonly employed in QD fabrication.

## 4. Experimental Section/Methods

*Chemicals:* Sulfur powder (S, 99.98%), lithium aluminum hydride powder (LiAlH$_4$, 95%), lead (II) oxide (PbO, 99.999%), 1-octadecene (ODE, technical grade, 90 %), bis(trimethylsilyl) sulfide ((TMS)$_2$S, synthesis grade), tris(pentafluorophenyl)borane (BCF, 95%), 2-methoxyethanol (99.8%, anhydrous), ammonium acetate (NH$_4$OAc, 99.99% trace metals basis), oleylamine (OlAm, 98%) were purchased from Sigma-Aldrich. Tetrahydrofuran (THF, 99.8%), diethyl ether (99.5%), sodium chloride (NaCl, 99.5%), sodium sulfate, (99%, anhydrous), acetone (99.8%), hexane (95%), chloroform (98.8%), acetonitrile (99.9%), methanol (99.8%), isopropanol (99.8%) were purchased from Fisher Chemical. Ethanol (99.5%, extra dry), N,N-dimethylformamide (DMF, 99.8%, extra dry) were purchased from Acros Organics. Bismuth (III) acetate (Bi(OAc)$_3$, 99%), n-Octane (99%, extra pure) was purchased from Thermo Scientific. 1,2-Ethanedithiol (EDT, 99%), 2-aminoethanol (>99.0%), lead (II) iodide (PbI$_2$, 99.999%, ultra dry), lead (II) bromide (PbBr$_2$, 99.999%, ultra dry), and n-butylamine (>99.0%) were purchased from TCI. Silver acetate (Ag(OAc), 99%, anhydrous), zinc acetate dihydrate (ACS reagent), tetramethylammonium iodide (TMAI, 99%) were purchased from Alfa Aesar. Stearoyl chloride (97%) was purchased from Fluorochem. Oleic acid (OlAc, 99%) was purchased from abcr. Poly[bis(4-phenyl)(2,4,6-trimethylphenyl)amine] (PTAA, M$_w$ 25000) was purchased from Ossila. Silver pellets (Ag, 99.99%) and gold pellets (Au, 99,99%) were purchased from Kurt J. Lesker. The ITO-covered glass substrates were acquired from Yingkou Shangneng Photoelectric material Co.,Ltd., based in Yingkou City (China).

*Synthesis of bis(stearoyl) sulfide (St$_2$S):* St$_2$S was synthesized according to the published method[47] with modifications. S powder (1.47 g, 46 mmol) was dissolved in 400 mL of THF at room temperature and under a nitrogen atmosphere. Then, LiAlH$_4$ (1.52 g, 40 mmol) was added to the flask in several portions under stirring. The mixture was stirred for 40 min, resulting in a grey LiAlHSH suspension. Then, stearoyl chloride (40.5 mL, 120 mmol) was added dropwise into the LiAlHSH suspension under vigorous stirring. The reaction mixture was stirred at room temperature for 4 h under a nitrogen atmosphere. The mixture was extracted with diethyl ether (700 mL) and washed 6 times with saturated NaCl solution. The organic layer was dried over sodium sulfate, filtered, and evaporated to dryness. The white powder was washed with isopropanol, recrystallized from diethyl ether, and dried under vacuum overnight.

*Synthesis of PbS QDs:* PbS QDs were synthesized following the hot-injection method with modifications.[29] PbO (446.4 mg, 2 mmol), OlAc (1.5 mL, 4.8 mmol) and 20 ml of ODE were loaded in a 50 mL flask and heated up to 100 °C and keep with this temperature for 2.5 h under vacuum to obtain a clear solution and remove water. Then, the system was filled with N$_2$, and the corresponding sulfur precursor was introduced into the reaction system:

<u>*i) Synthesis with St$_2$S.*</u> A warm solution (60 °C) of St$_2$S (567 mg, 1 mmol) in 1 mL of toluene was injected into the flask. After 5 seconds, 2 mL of OlAm was swiftly injected into the reaction system. The QDs grew for 10 seconds, and the reaction was quenched by rapid cooling with a water bath. The reaction mixture was stirred at room temperature for 20 min.

<u>*ii) Synthesis with TMS$_2$S*</u>. TMS$_2$S (210 µL, 1 mmol) diluted in 5 mL of ODE was swiftly injected into the flask. The heating mantle was removed, and the reaction mixture was allowed to cool down to room temperature for 20 min.

Then, PbS QDs were precipitated from the crude solution by adding acetone with subsequent centrifugation. The supernatant was discarded, and the precipitate was dispersed in 7 mL of hexane and stirred overnight. The solution was centrifuged for 3 min at 10000 rpm (10956 RCF). The precipitate was discarded, and acetone was added in portions until the solution became turbid. The QDs were precipitated with centrifugation and purified through precipitation and redispersion (2 times) using chloroform and acetonitrile as solvent and nonsolvent, respectively. Purified QDs were dispersed in octane for further use (optical density of 5 µL of this solution diluted with 2.5 mL of octane was 2.25 at 300 nm, which corresponds concentration of ≈ 50 mg/ml).

*Synthesis of AgBiS$_2$ QDs:* AgBiS$_2$ QDs were synthesized following the hot-injection method[27] with modifications. Ag(OAc) (133.5 mg, 0.8 mmol), Bi(OAc)$_3$ (386,1 mg, 1 mmol), OlAc (6 mL, 19.2 mmol) and 4 ml of ODE were loaded in a 50 mL flask and heated up to 100 °C for 2h under vacuum. Then, the system was filled with N$_2$ and the corresponding sulfur precursor was introduced into the reaction system:

<u>*i) Synthesis with St$_2$S*</u>. A warm solution (60 °C) of St$_2$S (567 mg, 1 mmol) in 1 mL of toluene was injected into the flask. The heating mantle was removed, and the reaction mixture was allowed to cool down to room temperature. The reaction mixture was stirred at room temperature for 60 min.

<u>*ii) Synthesis with (TMS)$_2$S*</u>. (TMS)$_2$S (210 µL, 1 mmol) diluted in 1 mL of ODE was swiftly injected into the flask. The heating mantle was removed and the reaction system was rapidly cooled down with a water bath. The reaction mixture was stirred at room temperature for 60 min.

Then, AgBiS$_2$ QDs were precipitated from the crude solution by the addition of acetone with subsequent centrifugation. The supernatant was discarded, and the precipitate was dispersed in 6 mL of hexane and stirred overnight. The solution was centrifuged for 3 min at 10000 rpm (10956 RCF). The precipitate was discarded, acetone was added in small portions until the solution became turbid and the QDs were precipitated with centrifugation.

In the case of AgBiS$_2$ QDs obtained with St$_2$S, the nanocrystals were additionally purified through precipitation and redispersion (4 times) using chloroform and acetonitrile as solvent and nonsolvent, respectively, to remove the stearic acid. In the case of AgBiS$_2$ QDs obtained with (TMS)$_2$S, the nanocrystals were additionally precipitated by adding ethanol, to remove residuals of unreacted salts.

Purified QDs were dispersed in toluene for further use (optical density of 10 µL of this solution diluted with 2.5 mL of anhydrous toluene was 2.5 at 300 nm, which corresponds to a concentration of 20 mg/ml).

*PbS QDs ligand exchange*: The DMF solution containing PbI$_2$ (0.1M), PbBr$_2$ (0.025M) and NH$_4$OAc (0.05M) was mixed with the solution of as-synthesized PbS QDs in octane (5 mg/mL) in a volume ratio of 1 to 1. The two-phase mixture was shaken vigorously until complete phase transfer of PbS QDs from octane to DMF. The DMF phase was collected and washed twice with pure n-octane. The lead halide capped QDs were precipitated from DMF by adding toluene (dropwise until the dispersion becomes turbid), followed by centrifugation (4000 rpm). The obtained precipitate was dried in a vacuum oven at 10 mbar and 40°C for 10 minutes and redispersed in n-butylamine at the desired concentration (usually 400 mg/mL).

*PbS QDs device fabrication:* Patterned ITO (indium tin oxide) glass substrates were sonicated sequentially in acetone and isopropanol for 20 minutes followed by an oxygen plasma treatment at 100 W, 0.4 mbar for 10 min (Diener Zepto). The ZnO precursor solution was prepared by dissolving zinc acetate (605.4 mg) in 2-methoxyethanol (6 mL) and 2-aminoethanol (181 µL). The mixture was stirred at 70 °C for 12 h. The ZnO precursor solution was then spin-coated at 3000 rpm for 15 s in two layers and annealed at 100°C (first layer) and 300°C (second layer) for 10 minutes each. Then, one layer of the halide capped PbS QDs is spun on the substrate at 2500 rpm for 15 s and annealed at 70°C for 5 minutes. Two layers of PbS-EDT were spin-coated in the air using already published methods.[63] Finally, 80 nm Au are thermally evaporated on the substrate, in a chamber at a pressure of 1e$^{-6}$ mbar.

*AgBiS$_2$ QDs device fabrication:* The ITO substrates were prepared as described above. The ZnO precursor solution was prepared by dissolving zinc acetate (660 mg) in ethanol (6 mL) and 2-aminoethanol (181 µL). The mixture was stirred at 80 °C for 3 h. The ZnO precursor solution was then spin-coated at 3000 rpm for 30 s in two layers and annealed at 200°C for 30 minutes each. AgBiS$_2$ QDs in toluene were dynamically spun at 2000 rpm and directly treated with TMAI in MeOH (1 mg/mL) two times for 20 sec, rinsed twice with MeOH and toluene. Three layers of AgBiS$_2$ were prepared in this fashion, annealed afterwards in ambient at 100 °C, resulting in a film thickness of ca. 35 nm. Solutions of PTAA (6 mg/mL) and BCF (2.4 mg/mL) in toluene were prepared and mixed in a volume ratio of (1:1). The resulting, intensively red solution was dynamically spun at 2000 rpm for 30 sec in a nitrogen filled glovebox and annealed at 100 °C. Finally, 80 nm of silver were thermally evaporated on the substrate through a shadow mask at a chamber pressure of $10^{-6}$ mbar.

*Transmission Electron Microscopy (TEM):* Images were acquired on a Jeol JEM F200 operated at 200 kV in transmission mode and images were taken on a Gatan 4k video camera. Samples were freshly drop-casted from a dilute QDs solution in toluene onto TEM carbon/copper grids and then measured directly on the microscope.

*UV-vis Spectroscopy:* Absorbance measurements were carried out at room temperature using a Jasco V-770 UV-vis spectrometer, from their diluted solutions in 1 cm-path quartz cuvettes.

*Photoluminescence measurements*: PL measurements were conducted using a pulsed 405 nm laser as excitation source (LDH-IB-405-B, PicoQuant) and a QDs dispersion in a sealed glass cuvette. Emitted light collected by an objective and separated by dichroic mirrors and long-filters the laser light and guided through a spectrograph (Kymera 328i, Andor) via a grating to a peltier-cooled silicon camera (iDus 420, Andor, calibrated by an external lamp source) for collection of a spectrum (integration time 50 ms) or to an avalanche detector (PDM, Micro Photon Devices) for time-correlated single photon counting measurements (TCSPC) using a PicoHarp 300, PicoQuant as central electronics for timing. Measurements were collected using an average power of 2.47 µW laser power and a repetition rate of 200 kHz.

*X-Ray Diffraction (XRD):* X-Ray spectra are collected in a Bruker D8 Discover diffractometer equipped with a 1.6 kW Cu X-ray filament ($\lambda$ = 1.5 Å) and a 2 mm slit. The scans were performed using a 1D detector in the θ/2θ mode. Samples for XRD are prepared by drop-casting the QD solution onto a clean glass substrate, with a thickness around 1 μm.

*X-ray Photoemission Spectroscopy (XPS):* XPS measurements were carried out in an ultrahigh vacuum chamber (ESCALAB 250Xi by Thermo Scientific, base pressure: $2 \times 10^{-10}$ mbar) using an XR6 monochromated Al Kα source (hν = 1486.6 eV) and a pass energy of 20 eV.

*Solar Cell Characterization:* Devices are measured under the illumination of a solar simulator (Abet, Class A+++) and using a Keithley SMU 2450 for precise current-voltage scans (−0.1 to 0.7 V in 0.025 V steps). EQE spectra were measured with a self-built setup. A halogen lamp passed through a monochromator was used to generate a current in the solar cells, which was measured with a Keithley SMU. This current was corrected for the irradiance of the lamp using a calibrated silicon diode and the efficiency to generate an electron for every irradiated photon was calculated.

**Supporting Information**

Supporting Information is available from the Wiley Online Library or from the author.

**Acknowledgements**

M.A-S. and A.P. contributed equally to this work. A.P., A.S. and F.P. thank the German Bundesministerium für Bildung und Forschung (BMBF) for funding (Förderkennzeichen 03XP0422 "GreenDots"). This project has received funding from the European Research Council (ERC) under the European Union's Horizon 2020 research and innovation programme (ERC Grant Agreement n° 714067, ENERGYMAPS). We thank the Faculty for Chemistry and Food Chemistry at TU Dresden, in particular Prof. B. Plietker, for access to NMR facilities. M.A-S. and A.P. would like to thank the Dresden Center for Nanoanalysis (DCN), for access to the TEM microscope.